\documentstyle[12pt,rotate]{article}
\newcommand{\be}{\begin{equation}}
\newcommand{\ee}{\end{equation}}
\newcommand{\bea}{\begin{eqnarray}}
\newcommand{\eea}{\end{eqnarray}}
\newcommand{\bd}{\begin{displaymath}}
\newcommand{\ed}{\end{displaymath}}

\begin{document}
\bibliographystyle{physics}
\renewcommand{\thefootnote}{\fnsymbol{footnote}}

\author{
Dongsheng Du${}^{a,b}$~~~ Hongying Jin${}^b$~~~  Yadong Yang${}^{a,b,c}$
\thanks{Email: duds@bepc3.ihep.ac.cn,~~jhy@hptc5.ihep.ac.cn, 
~yangyd@bepc3.ihep.ac.cn}\\
{\small\sl ${}^{a}$ CCAST (World Laboratory), P.O.Box 8730, Beijing 
100080, China}\\
{\small\sl ${}^{b}$ Institute of High Energy Physics, Academia Sinica,
P.O.Box 918(4), Beijing 100039, China\thanks{Mailing address} }\\
{\small\sl ${}^{c}$ Physics Department of Henan Normal University,
Xingxiang,
Henan, 453002, China }
}
\date{}
\title{
{\large\sf
\rightline{BIHEP-Th/97-009}
}
\vspace{3cm}
\bigskip
\bigskip
{\LARGE\sf Nonfactorization Effects in  Inclusive Decay $B\rightarrow
J/\Psi +X_{s}$ }
 } 
\maketitle
\begin{abstract}
\noindent
We discuss the nonfactorization effects in $B\rightarrow J/\Psi+X_{s}$, 
which is similar to the nonperturbative effect found 
by Voloshin in the  decay $B\rightarrow \gamma +X_{s}$. 
The QCD sum rule has been used to estimate the hadron 
matrix elements.  We find that the correction from this effect is very large
and  the large discrepancy between the theory and
the experimental data can be  reduced considerably.

\end{abstract}
\vspace{1.5cm}
{\bf PACS  numbers 13.25.Hw 12.28.Lg } 

\newpage

It is amazing  that there is a large discrepancy between
the theoretical predictions, which is based on the operator 
product expansion (OPE) with assumption of 
the factorization, and experimental data in the
process  $B\rightarrow J/\Psi +X_{s}$.
The theoretical prediction   of the branching  ratio  
 $B\rightarrow J/\Psi +X_{s}$ is $0.23\%$, which is only 
 one-third of the  experiment data $(0.8\pm0.08)\%$\cite{data}. 
 In order to resolve 
this puzzle, some scenarios have been suggested \cite{sum}. Besides 
the possible existence of new charmonium states above the ${\bar D}D$ 
threshold, nonfactorization effects have been widely considered.
In the later case, color-octet 
effect seems to be the most reasonable source that may enhance the 
theoretical calculation.
Color-octet mechanism was first used in  calculating 
cross section of ${\bar c}c$
production, where the theoretical  prediction  based on color-singlet 
mechanism is also much smaller than the experimental  data. In this
mechanism, the gluon parton 
has been taken account  so that the color-octet operators responsible
for $b\rightarrow{\bar c}c +q$  may give a non-zero contribution to the
decay mentioned above. However, the most important parameter, 
the matrix element of color-octet operators, is not calculable so far. 
It needs future determination  both from  various  
experiments and theoretical estimations. 
  
Recently, a nonperturbative correction of order  
of $O(\Lambda^2_{QCD}/m^2_c)$ to $B\rightarrow\gamma+X_{s}$ 
is first discussed by Voloshin\cite{Voloshin} ( a simlar correction 
is also independently discussed by \cite{Ruckl} ) 
which arises from   the
contribution of the gluon-photon penguin graph shown in Fig.1, and then  
 discussed  by Grant $et~al$ , Legeti
$et~al$\cite{wise} and Buchalla $et~al$\cite{Buch} .  It is  obvious  that
this effect also exits in $B\rightarrow J/\Psi+X_{s}$ as 
nonfactorization effects. 
Although this correction is not large in $B\rightarrow\gamma +X_{s}$,
one can expect a considerable contribution from this mechanism in
 $B\rightarrow J/\Psi+X_{s}$ due to 
the large ratio of the Wilson Coefficients between color-octet operator
and color-singlet operator responsible for $B\rightarrow
J/\Psi+X_{s}$.
QCD sum rule\cite{shifman} so far 
is a powerful  tool to deal with hadron matrix elements. Although there 
is an up to 30 percent uncertainty in this method, we can still  
give a useful estimation of the nonfactorization effect mentioned
above. \\

The effective weak interaction Hamiltonian at a scale $\mu$ is given by 
\be
H_{eff}=-{{4G_F}\over{\sqrt{2}}}V_{cb}V^*_{cs}\sum^2_{i=1}C_i(\mu)O_i(\mu),
\ee
in the conventional notation,
$O_2 =({\bar s}_{L}\gamma_{\mu}\frac{\lambda^a}{2} b_{L})
({\bar c}_{L}\gamma^{\mu} \frac{\lambda^a}{2} c_{L})$, 
$O_1$ only differs from $O_2$ in the way of color indices contraction, 
and the former  is referred as color-octet operator.
On the factorization assumption, the matrix element of $O_1$
dominates   the decay $B\rightarrow J/\Psi X_{s}$
while $ O_2$ gives zero.  

The computation of the inclusive $B\rightarrow J/\Psi X_{s}$ decay 
rate from the $O_1$ contribution is performed by calculating the
correlation function 
\be 
T=i\int d^4x e^{-iq\cdot x}\langle
B(v)|T[j^\dagger_\mu(x)j_\nu(0)]|B(v)\rangle
(-g^{\mu\nu}+{{q^\mu q^\nu}\over{M_\Psi^2}}), 
\ee
where $j_\mu={\bar s}_L\gamma_\mu b_L$. The corresponding decay rate  
is given by 
\be
{{d\Gamma}\over{P_{J/\Psi}dE_{J/\Psi}}}=
{{G_F^2f^2_{J/\Psi}|V^*_{ts}V_{tb}|^2 C_1^2}\over
{2M_B\pi^2}}ImT ,
\label{fa}
\ee
where $f_{J/\Psi}$ is defined by   
$\langle 0|{\bar c}\gamma_\mu c|J/\Psi(p,\epsilon)\rangle=i
f_{J/\Psi}\epsilon_\mu$.

At the leading order in the operator product expansion (OPE), 
ImT has the form
\be
ImT=\pi\delta(m_b^2+M^2_{J/\Psi}-2m_bE_{J/\Psi})M_B
{{2E_{J/\Psi}^2m_b-3E_{J/\Psi}M^2_{J/\Psi}+M^2_{J/\Psi}m_b}\over
{M^2_{J/\Psi}}}.   
\ee
It is well known that the next to leading order contribution  in OPE is 
suppressed by $O(\Lambda^2_{QCD}/m^2_b)$. Therefore, on the 
factorization approximation, 
(\ref{fa}) almost gives the total decay rate of  
$B\rightarrow J/\Psi X_{s}$. 
The main non-factorization contribution arises from the
interference terms of $ O_1$ and $O_2$ as shown in   
 Fig.2, which is of order of $\Lambda^2_{QCD}/m^2_c$\cite{Voloshin}.
The coefficient ratio of operator $O_2$ over 
$O_1$ is $ C_2$/$C_1 $ 
 $\sim$ 20. Therefore, it is reasonable to expect a large
enhancement when  contribution of  Fig.2 is taken into account.\\ 
The decay rate from Fig.2 is obtained as 
\be
d\Gamma
=
{{2G_F^2|V^*_{ts}V_{tb}|^2{C}_1*C_2}\over
{2M_B}}
[ImI_{\nu\mu\beta\alpha} T^{\nu\mu\beta\alpha}
+ImI'_{\nu\mu\beta\alpha}{T'}^{\nu\mu\beta\alpha}]
{{dp^4}\over{(2\pi)^4}}.
\ee
where $I_{\nu\mu\beta\alpha}$  and $T^{\nu\mu\beta\alpha}$
are defined as 
\be
\begin{array}{lll}
I_{\nu\mu\beta\alpha}&=&\int d^4x 
{x_\beta\over 16}\pi\delta(p^2-M^2_{J/\Psi})
\langle 0|j^\nu|J/\Psi(p)\rangle
\langle J/\Psi(p)|j^a_\alpha(x)j^{a5}_\mu(0)|0\rangle\\
&=&Im\int d^4x {x_\beta\over 16}d^4y e^{ip\cdot y}
i\langle 0|Tj_\nu(y) j^a_\alpha(x)j^{a5}_\mu(0)|0\rangle\\
&=&Im -i{\partial\over{\partial q^\beta}}\{\int d^4x {1\over 16}d^4y
e^{iq\cdot x}e^{ip\cdot y}
i\langle 0|Tj_\nu(y) j^a_\alpha(x)j^{a5}_\mu(0)|0\rangle\}|_{q=0}\\
&=&Im K_{\nu\mu\beta\alpha}(p)~,\\

T^{\nu\mu\beta\alpha}&=&\int d^4x e^{-ip\cdot x} 
i\langle B|{\bar b}_L(x)\gamma^\nu S_s(x,0)\gamma^\mu
(-ig_sG^{\beta\alpha})
b_L(0)|B\rangle~, \\
I'_{\nu\mu\beta\alpha}
&=&\pi\delta(p^2-M_{J/\Psi})\langle
0|j_\nu|J/\Psi(p)\rangle\langle J/\Psi(p)|
{\bar c}\gamma_\mu (-ig_sG_{\beta\alpha})c|0\rangle\\
&=&Im\int d^4x e^{ip\cdot x}
i\langle 0|j_\nu(x) {\bar
c}(0)\gamma_\mu (-ig_sG_{\beta\alpha})c(0)|\rangle\\
&=&Im K'_{\nu\mu\beta\alpha}(p)~,\\
{T'}^{\nu\mu\beta\alpha}&=&\int d^4x d^4y e^{-ip\cdot x}
i{x^\beta\over 12}\langle B|{\bar b}_L(y)\gamma^\nu 
S_s(y,x)\gamma^\alpha S_s(x,0)
\gamma^\nu b_L(0)|B\rangle~\\.
\end{array}
\label{it}
\ee

In equation (\ref{it}), we have assumed that other resonances contributions
to $I$ and $I'$ are small compared with $J/\Psi$ . 
$S_s(x,y)$ is the propagator of $s$ quark, 
$j_\mu={\bar c}\gamma_\mu c$,
$j^a_\mu={\bar c}\gamma_\mu{\lambda^a\over 2}c$ and 
$j^{a5}_\mu={\bar c}\gamma_5\gamma_\mu{\lambda^a\over 2}c$. 
We  only keep  the leading order terms in OPE.  $Im$ in the definition 
of $I$ and $I'$ should be understood as an operator which selects the
imaginary part of the scalar form factors which factorize the followed
matrix element.\\  
In this paper, $I$ and $I'$ are estimated  by QCD sum rule. The
matrix elements $K_{\nu\mu\beta\alpha}$ 
$K'_{\nu\mu\beta\alpha}$
in  equation (\ref{it}) are obtained 
by calculating the diagrams shown in Fig.3 and Fig.4,
\be
\begin{array}{lll}
K_{\nu\mu\beta\alpha}&=&K^0_{\nu\mu\beta\alpha}
+2K^{a}_{\nu\mu\beta\alpha}+2K^{b}_{\nu\mu\beta\alpha}
+2K^{c}_{\nu\mu\beta\alpha},\\
K'_{\nu\mu\beta\alpha}&=&2K^{d}_{\nu\mu\beta\alpha}.
\end{array}
\ee
where $K^0$ is from  Fig.3 and others are from Fig.4 and correspond
to the diagrams with the indices $a,~b,~c$ and $d$ respectively.

In a conventional way, the fixed point gauge  $x\cdot A(x)=0$ is
taken. At the first order of the expansion in term of 
$x^\mu$ is,
\be 
A_\mu(k)=\displaystyle{{{-i(2\pi)^4}\over{2}}G_{\nu\mu}(0)
{\partial\over{\partial k^\nu}}\delta^4(k)}.
\ee
 We obtain 
\be
\begin{array}{lll}
K^0_{\nu\mu\beta\alpha} 
{\leftskip 3cm }=i4\displaystyle{\int \frac{d^Dk}{(2\pi)^D}{\partial\over{\partial
q^\beta}}}&~&\\
\displaystyle{\
Tr\{\frac{i}{{\hat k}-m_c}  \gamma_\alpha \frac{i}
{{\hat k}-{\hat q}-m_c}
\gamma_\nu\frac{i}{\hat k-\hat p-\hat q-m_c}\gamma_5\gamma_\mu 
+\frac{i}{{\hat k}-m_c}\gamma_\mu\frac{i}{{\hat k}-{\hat q}-m_c}
\gamma_5\gamma_\mu\frac{i}{\hat k+\hat p-m_c}\gamma_\nu \}|_{q=0}
 },&~&\\

K^{a}_{\nu\mu\beta\alpha} 
=\displaystyle{-\frac{i}{96}(
\delta_{\rho\sigma}\delta_{\rho'\sigma'} 
-\delta_{\rho\sigma'}\delta_{\sigma\rho'})\langle g^2_sG^2\rangle
\frac{1}{4}Tr\{\frac{\lambda^c}{2}\frac{\lambda^a}{2}\frac{\lambda^c}{2}
\frac{\lambda^a}{2}\}
\int \frac{d^Dk}{(2\pi)^D}
\frac{\partial}{\partial
q^\beta}\frac{\partial}{\partial k^\rho_1}\frac{\partial}{\partial 
k^\sigma_2}}&~&\\
\displaystyle{
Tr\{\frac{i}{\hat k-m_c}
\gamma_\alpha \frac{i}{\hat k-\hat q-m_c}
\gamma_{\sigma'}\frac{i}{\hat k-\hat q+{\hat k}_2-m_c}
\gamma_5\gamma_\mu
\frac{i}{\hat k+\hat p-{\hat k}_1-m_c}\gamma_\nu
\frac{i}{\hat k-{\hat k}_1-m_c}\gamma_{\rho'}\}|_{q=k_1=k_2=0}},\\

K^{b}_{\nu\mu\beta\alpha} 
=\displaystyle{-\frac{i}{96}(
\delta_{\rho\sigma}\delta_{\rho'\sigma'} 
-\delta_{\rho\sigma'}\delta_{\sigma\rho'})\langle g^2_sG^2\rangle
\frac{1}{4}Tr\{\frac{\lambda^c}{2}\frac{\lambda^a}{2}\frac{\lambda^c}{2} 
\frac{\lambda^a}{2}\}
\int \frac{d^Dk}{(2\pi)^D}
\frac{\partial}{\partial
q^\beta}\frac{\partial}{\partial k^\rho_1}\frac{\partial}{\partial 
k^\sigma_2}}&~&\\
\displaystyle{
Tr\{\frac{i}{{\hat k}-m_c}
\gamma_\alpha \frac{i}{\hat k-\hat q-m_c}
\gamma_{\rho'}\frac{i}{\hat k-\hat q+{\hat k}_1-m_c}
\gamma_5\gamma_\mu
\frac{i}{\hat k+\hat p-{\hat k}_2-m_c}\gamma_{\sigma'}
\frac{i}{\hat k+\hat p-m_c}\gamma_{\nu}\}|_{q=k_1=k_2=0}},\\

K^{c}_{\nu\mu\beta\alpha}=\displaystyle{-\frac{i}{96}(
\delta_{\rho\sigma}\delta_{\rho'\sigma'} 
-\delta_{\rho\sigma'}\delta_{\sigma\rho'})\langle g^2_sG^2\rangle
\frac{1}{4}Tr\{\frac{\lambda^c}{2}\frac{\lambda^c}{2}\frac{\lambda^a}{2} 
\frac{\lambda^a}{2}\}
\int \frac{d^Dk}{(2\pi)^D}
\frac{\partial}{\partial
q^\beta}\frac{\partial}{\partial k^\rho_1}\frac{\partial}{\partial 
k^\sigma_2}}&~&\\
\displaystyle{
Tr\{\frac{i}{{\hat k}-m_c}
\gamma_\alpha \frac{i}{\hat k-\hat q-m_c}
\gamma_5\gamma_\mu\frac{i}{\hat k+\hat p-{\hat k}_1
-{\hat k}_2-m_c}
\gamma_{\sigma'}
\frac{i}{\hat k+\hat p-{\hat k}_1-m_c}\gamma_\nu
\frac{i}{\hat k-{\hat k}_1-m_c}\gamma_{\rho'}\}|_{q=k_1=k_2=0}
},&~&\\

K^d{\nu\mu\beta\alpha}=\displaystyle{-\frac{1}{48}(
\delta_{\sigma\beta}\delta_{\rho\alpha}
-\delta_{\sigma\alpha}\delta_{\rho\beta})\langle g^2_sG^2\rangle
}
\displaystyle{\int \frac{d^Dk}{(2\pi)^D}
{\partial\over \partial q^\sigma}
Tr\{\frac{i}{{\hat k}-m_c}\gamma_\rho\frac{i}{\hat k-\hat q-m_c}
\gamma_\mu\frac{i}{\hat k-\hat p-m_c}\}|_{q=0} }.&~&\\
\end{array}
\label{I} 
\ee

The computation of (\ref{I}) is tedious, we will not display the
procedure. Instead, we  give the final result at the end of
this paper. 

Using the dispersion relation 
\bea
K_{\nu\mu\beta\alpha}(p)={1\over \pi}\int {{Im
K_{\nu\mu\beta\alpha}(s)}\over 
{s-p^2}}ds,\\ 
K'_{\nu\mu\beta\alpha}(p)={1\over \pi}\int {{Im
K'_{\nu\mu\beta\alpha}(s)}\over 
{s-p^2}}ds,
\eea 
 and the fact that $I_{\nu\mu\beta\alpha}$ and $I'_{\nu\mu\beta\alpha}$ 
are  proportional to $\delta(p^2-M_{J/\Psi})$, we get two sum rules 
\bea
I_{\nu\mu\beta\alpha}(p)
=\pi\delta(p^2-M^2_{J/\Psi})M^4_{J/\Psi}
[{d\over{dP^2}}K_{\nu\mu\beta\alpha}(p)|_{p^2=0}],\\
I'_{\nu\mu\beta\alpha}(p)
=\pi\delta(p^2-M^2_{J/\Psi})M^4_{J/\Psi}
[{d\over{dP^2}}K'_{\nu\mu\beta\alpha}(p)|_{p^2=0}],
\eea
where $P^2=-p^2$. 
 ${d\over{dP^2}}K_{\nu\mu\beta\alpha}(p)|_{p^2=0}$ must be
understood that only  derivatives of the scalar form factors of the
matrix element are set at $p^2=0$. To choose $p^2=0$ is for the sake of
convenience. In general, we can set $p^2$ at any point in the range
$-p^2+m_c^2>>\Lambda^2_{QCD}$ 
to get a sum rule.  In order to get rid of the dependence on the 
subtraction in the loop calculation, we have used  the sum rules for 
derivatives of $K_{\nu\mu\beta\alpha}$ and $K'_{\nu\mu\beta\alpha}$ 
instead of themselves.

The calculations of  $T^{\nu\mu\beta\alpha}$ and 
$T'^{\nu\mu\beta\alpha}$ are performed, in which we need  
the identities 
\bea 
{1\over{2M_B}}
\langle B(v)|{\bar b}\Gamma g_s G_{\alpha\beta}b|B(v)\rangle 
=-{\mu^2_g\over{24}}
Tr\{\Gamma(1+{\hat v})\sigma_{\alpha\beta}(1+{\hat
v})\},\\
\langle B(v)|{\bar b}\Gamma b|B(v)\rangle 
={M_B\over 4}
Tr\{(1+{\hat v})\Gamma(1+{\hat v})\},
\eea
where $\Gamma$ is any kind of Dirac structure and $\mu^2_g$ is the value
of the strength of the chromomagnetic 
interaction of the b-quark inside in B hadron,
\be
\mu^2_g={1\over {2M_B}}\langle B(v)|{\bar
b}\sigma_{\alpha\beta}G^{\alpha\beta}_a
{\lambda^a\over 2}b|B(v)\rangle={3\over 4}({M^*_B}^2-M^2_B)\approx
0.4
GeV^2.
\ee
  
After a tedious calculation, we arrive at 
\be
\begin{array}{lll}
ImI_{\nu\mu\alpha\beta}T^{\nu\mu\alpha\beta}
&=&\displaystyle{{\mu^2_gM^4_{J/\Psi}M_B\over 12}
\pi^2\delta(m_b^2+M^2_{J/\Psi}-2m_bE_{J/\Psi})
\delta(p^2-M^2_{J/\Psi})}\\
&&\displaystyle{({{4m_bE^2_{J/\Psi}-m_bm_c^2-3m^2_cE_{J/\Psi}}
\over{240\pi^2m_c^4}}-{{44m_bE^2_{J/\Psi}-33m_bm^2_c-11m_c^2E_{J/\Psi}}
\over{12096m_c^8}}{{\langle\alpha_sG^2\rangle}\over{24\pi}})}\\
ImI'_{\nu\mu\alpha\beta}{T'}^{\nu\mu\alpha\beta}
&=&\displaystyle{M^4_{J/\Psi}\pi^2\delta(m_b^2+M^2_{J/\Psi}-2m_bE_{J/\Psi})
\delta(p^2-M^2_{J/\Psi})}\\
&&\displaystyle{{m_b^2+61m_c^2-6m_bE_{J/\Psi}}\over{240m_c^4}}
{{\langle\alpha_s G^2\rangle}\over{6\pi}}.
\label{re}
\end{array}
\ee
In the second equation of (\ref{re}), we have already used partial 
integration. Using the standard numerical values 
\be
\begin{array}{ll} 
\langle\alpha_s G^2\rangle=0.04GeV^4, ~~~~~m_c=1.3GeV, &m_b=4.5GeV\\
C_{1}(m_b)=(2C_{+}(m_b)-C_{-}(m_b))/3=0.133,&
C_{2}(m_b)=C_{+}(m_b)+C_{-}(m_b)=2.21,~\\  
f_{J/\Psi}=0.38GeV,&~~~~~~~~~~~~~~~~~~~~~~~~~~~~~~~~~~~      
\end{array}
\ee
we obtain the nonfactorization contribution arisen from color-octet 
amplitude as large as
\be
{{\delta\Gamma(B\rightarrow J/\Psi+X_s)}\approx 
1.25{\Gamma_0 (B\rightarrow J/\Psi+X_s)}},
\ee
where $\Gamma_0$ is  the result with factorization assumption.
This result is sensitive to the value of $m_c$, if we choose $m_c=1.5GeV$,
the ratio would be $\sim 0.7$. The non-perturbative effect  from the 
the gluon condense is much smaller compared with the perturbative 
diagram, so, the result is not sensitive to the parameter
$\langle\alpha_s G^2\rangle$.  
However,  
since the decay rate in (\ref{fa}) is  independent of  $m_c$, the
non-factorization contribution is very 
large any way. One may notice that we have only taken account of 
 the interference term shown in  fig.2. If the non-factorization
effects from pure $O_2$ contribution are  taken into account,
which are very complicated to be calculated and will be presented in 
a separate paper\cite{djy},
the enhancement will be hopeful to explain   the discrepancy between 
the theoretical predictions  and the experimental data. 
\noindent

{\large\bf Acknowledgment}

\noindent
This work is supported 
 in part by the National Natural Science Foundation and the 
Grant of State Commission of Science and Technology of China.
One of authors (Yang Y.D) thanks Dr.Isidori for pointing out 
a typo in eq.(5).

\newpage
\begin{center}
{\large Figure Captions}\\

\end{center}
\noindent Fig.1 Nonperturbative  effect in $B\rightarrow
X_s+\gamma$\\
Fig.2 Nonfactorization effect in $B\rightarrow X_s+J/\Psi$\\
Fig.3 The perturbative diagrams contribute to  
$K^{\nu\mu\beta\alpha}$\\
Fig.4 The gluon condense diagrams contribute to 
$K^{\nu\mu\beta\alpha}$ and ${K'}^{\nu\mu\beta\alpha}$

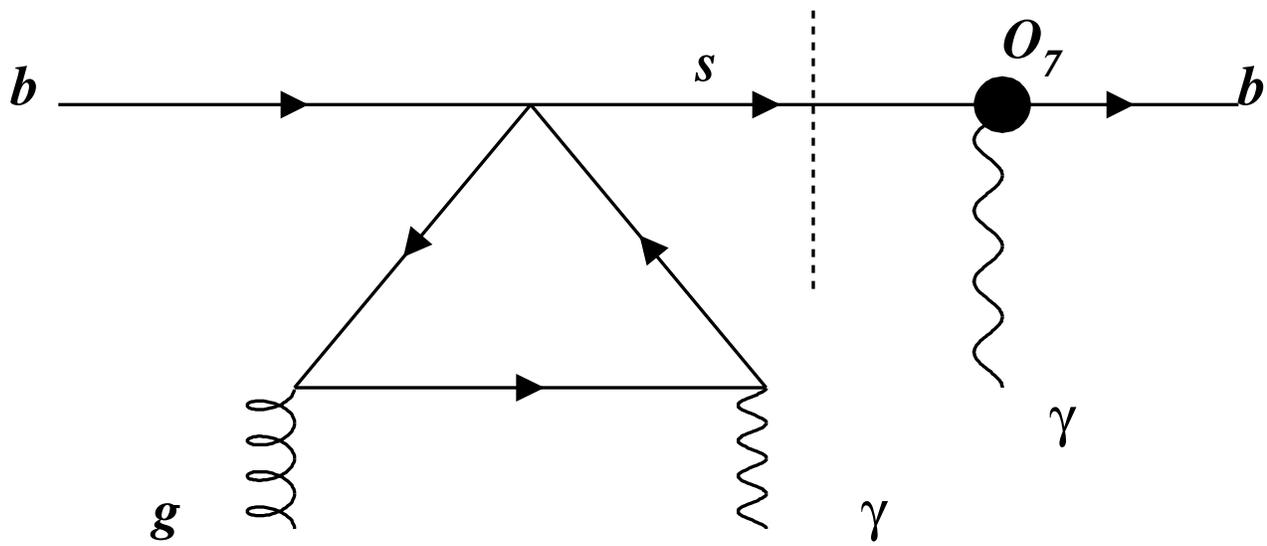
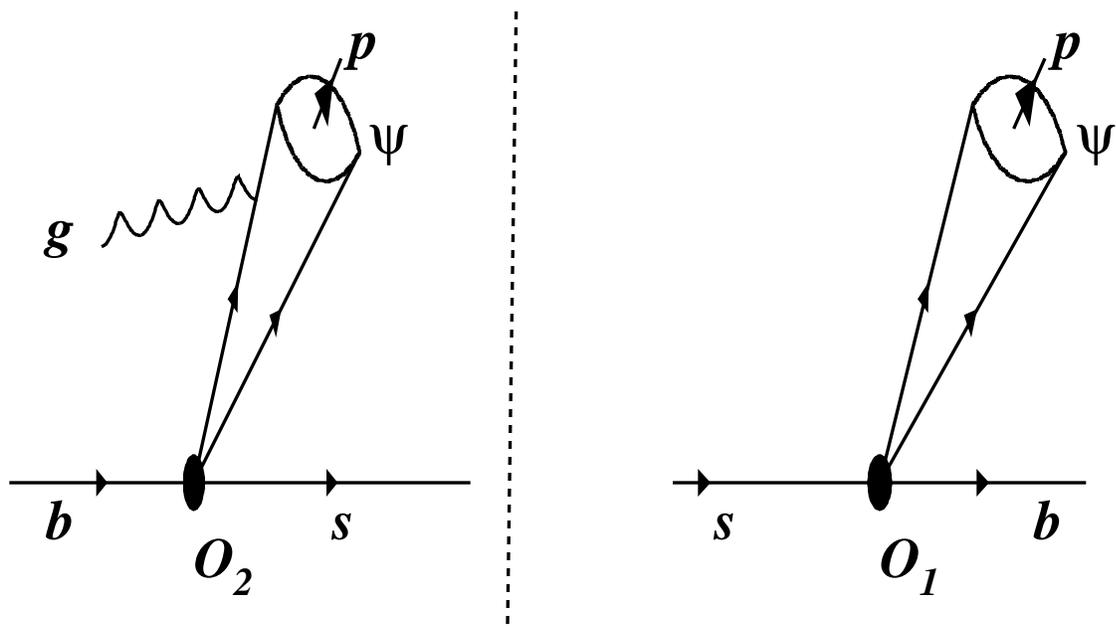
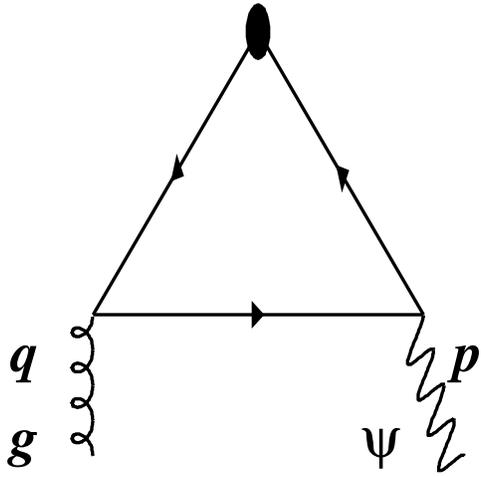
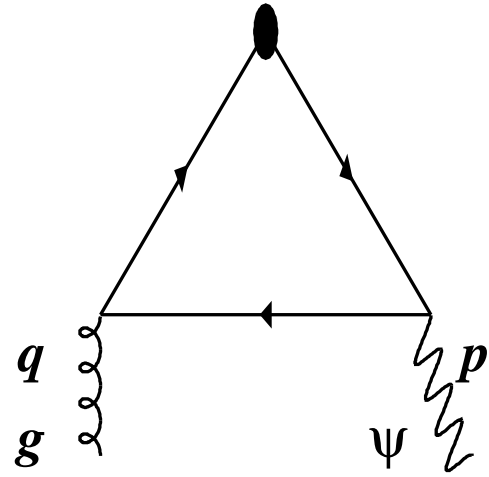
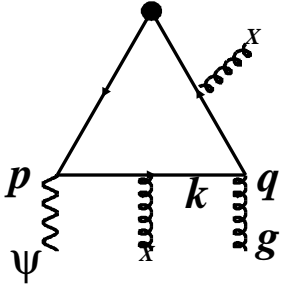
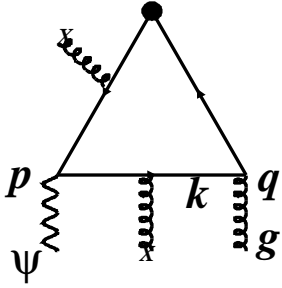
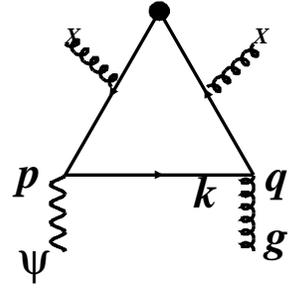
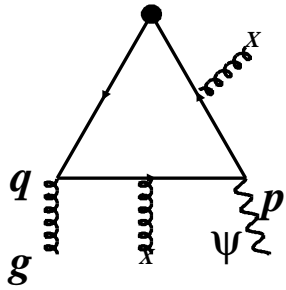
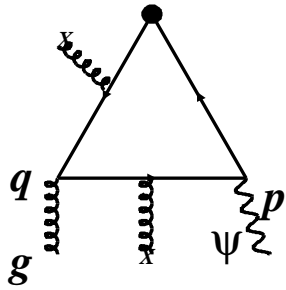
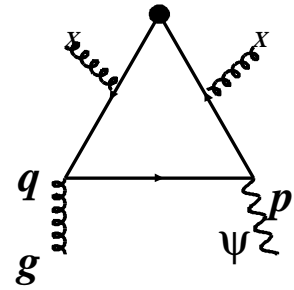
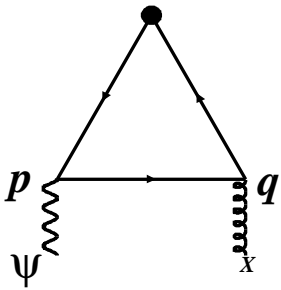
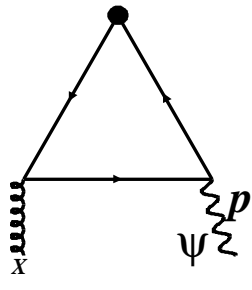
\end{document}